\RequirePackage{fix-cm}
\documentclass[smallextended]{svjour3}       
\smartqed  
\usepackage{graphicx}
\usepackage{hyperref}
\usepackage{url}
\usepackage{color}

\newcommand{\sax}{\emph{BeppoSAX}}

\journalname{Experimental Astronomy}

\begin{document} 

\title{The radiation environment in a Low Earth Orbit: \\ the case of \emph{BeppoSAX}}

\author{R.~Campana
        \and
        M.~Orlandini
        \and
        E.~Del~Monte
        \and
        M.~Feroci
        \and
        F.~Frontera
        }

\institute{ R. Campana, M. Orlandini, F. Frontera \at 
			INAF/IASF-Bologna, via Gobetti 101, I-40129 Bologna, Italy \\
			\email{riccardo.campana@inaf.it}             \\
			E. Del Monte, M. Feroci \at
            INAF/IAPS, via Fosso del Cavaliere 100, I-00133 Roma, Italy \\
            F. Frontera \at 
            Department of Physics and Earth Science, University of Ferrara, via Saragat 1, I-44100 	
            Ferrara, Italy
		  }

\date{Received: date / Accepted: date}

\maketitle
 
\begin{abstract}
Low-inclination, low altitude Earth orbits (LEO) are of increasing importance for astrophysical satellites, due to their low background environment. Here, the South Atlantic Anomaly (SAA) is the region with the highest amount of radiation.
We study the radiation environment in a LEO (500--600~km altitude, 4$^{\circ}$ inclination) through the  particle background measured by the Particle Monitor (PM) experiment onboard the \sax\ satellite, between 1996 and 2002.
Using time series of particle count rates measured by PM we construct intensity maps and derive SAA passage times and fluences.
The low-latitude SAA regions are found to have an intensity strongly decreasing with altitude and dependent on the magnetic rigidity. The SAA extent, westward drift and strength vs altitude is shown.

\end{abstract}


\section{Introduction}
Knowledge of the radiative environment surrounding a scientific satellite is of paramount importance in order to assess the behaviour of its instruments. High energy particles (e.g. 0.1--100 MeV protons) can penetrate the satellite shielding and, beside leaving their imprint as background events, can also damage the detectors and their electronics. 

Low Earth orbits (LEO), i.e. satellite orbits with an apogee lower than 1000--2000~km, are of great interest because they lie mostly outside the high-energy particle environment of the radiation belts.
However, the South Atlantic Geomagnetic Anomaly (SAGA or SAA) is a region roughly between the coasts of Brazil and South Africa where, due to the inclination and offset of the geomagnetic field axis with respect to the spin axis, the local magnetic field is weakened and the radiation belt edge reaches much lower altitudes. Here, the high energy trapped proton flux increases by several orders of magnitude.
Scientific instrumentation onboard satellites is usually switched off during SAA passages, due to the much higher background level. High energy protons in the SAA can also contribute to the radio-activation of materials surrounding the detector.

The internal dynamics of the geomagnetic field are complex \cite{wardinski08}, with secular variations (e.g. leading to a westward drift of the SAA \cite{pinto92}) superimposed to impulsive changes in the field configuration (geomagnetic jerks, in which the secular variation suddenly changes slope \cite{demichelis05}).
A characterisation and monitoring of the SAA environment, and generally speaking of the radiative environment in a low-Earth orbit, is therefore of particular importance to the design and operations of present and future instrumentation, especially for high energy astrophysics observations (e.g. LOFT \cite{feroci12}). 

The current models of particle fluxes (e.g. AP8/9, \cite{sawyer76,vette91,ginet13}) are usually based on large samples of observations performed by instruments in high-inclination, highly elliptic orbits. Only few instruments (in particular RXTE, \cite{furst09}) have observed the SAA environment over a long, continuous time basis, especially at low altitudes (below $\sim$800 km) and low inclination. 

In this paper we show the high-energy radiation environment measured almost uninterruptedly by the \sax\ mission during 1996--2002 (covering about half of the 24th solar cycle), on a low-inclination ($\sim$4$^{\circ}$), 500--600 km altitude orbit.
In particular, we independently confirm the main findings of RXTE over a similar epoch and altitude span (but at a different orbital inclination, $\sim$23$^{\circ}$), i.e. the rapid decrease of flux with respect to the altitude, and the westward drift of the longitude of maximum emission \cite{furst09}.
We also find a dependence of the SAA count rates on the local magnetic parameters, such as the cutoff rigidity, and on the solar cycle phase.

This paper is structured as follows. In Section \ref{saxmission} we briefly describe the \sax\ mission and its Particle Monitor instrument. In Section \ref{pm} the Particle Monitor data is shown, and in Section \ref{conclusions} we draw our conclusions.

\section{The BeppoSAX mission and the Particle Monitor}\label{saxmission}
\begin{figure*}[htbp]
\centering
\includegraphics[width=\textwidth]{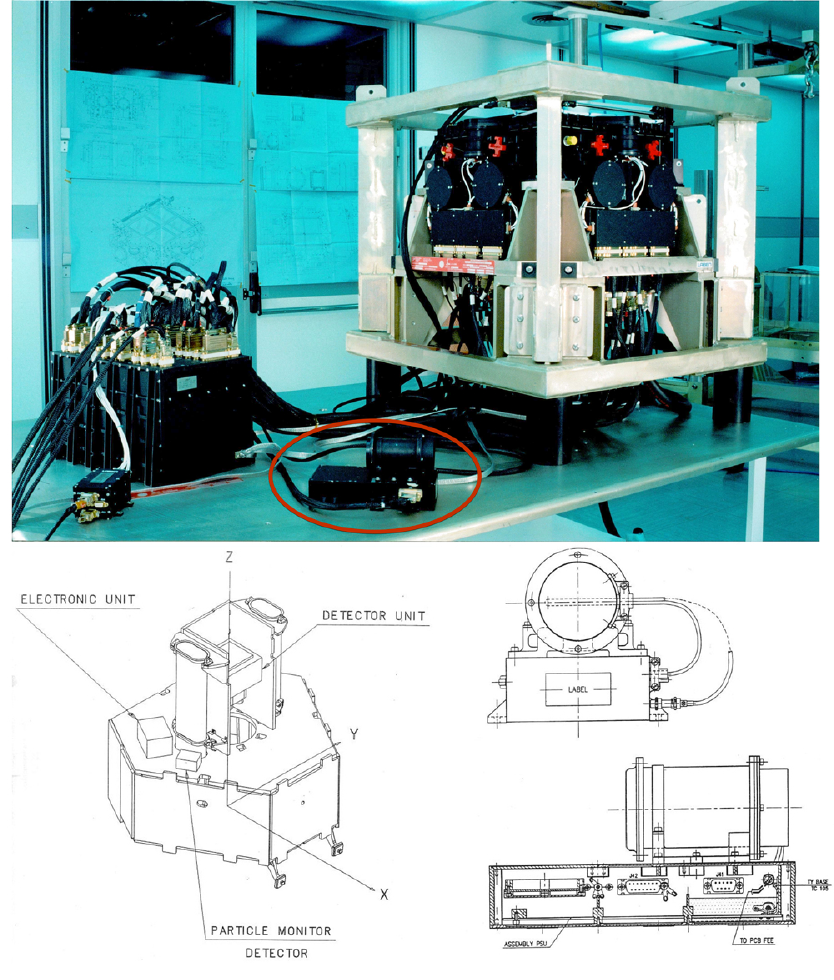}
\caption{ \emph{Upper panel:} The Particle Monitor (circled) before integration. \emph{Lower left panel:} The PM location with respect to PDS and the other instrumentation onboard \sax{}. \emph{Lower right panel:} Technical drawing of the PM tube and front-end electronics unit.}
\label{fig_saxpm}
\end{figure*}

\begin{figure}[htbp]
\centering
\includegraphics[width=0.9\textwidth]{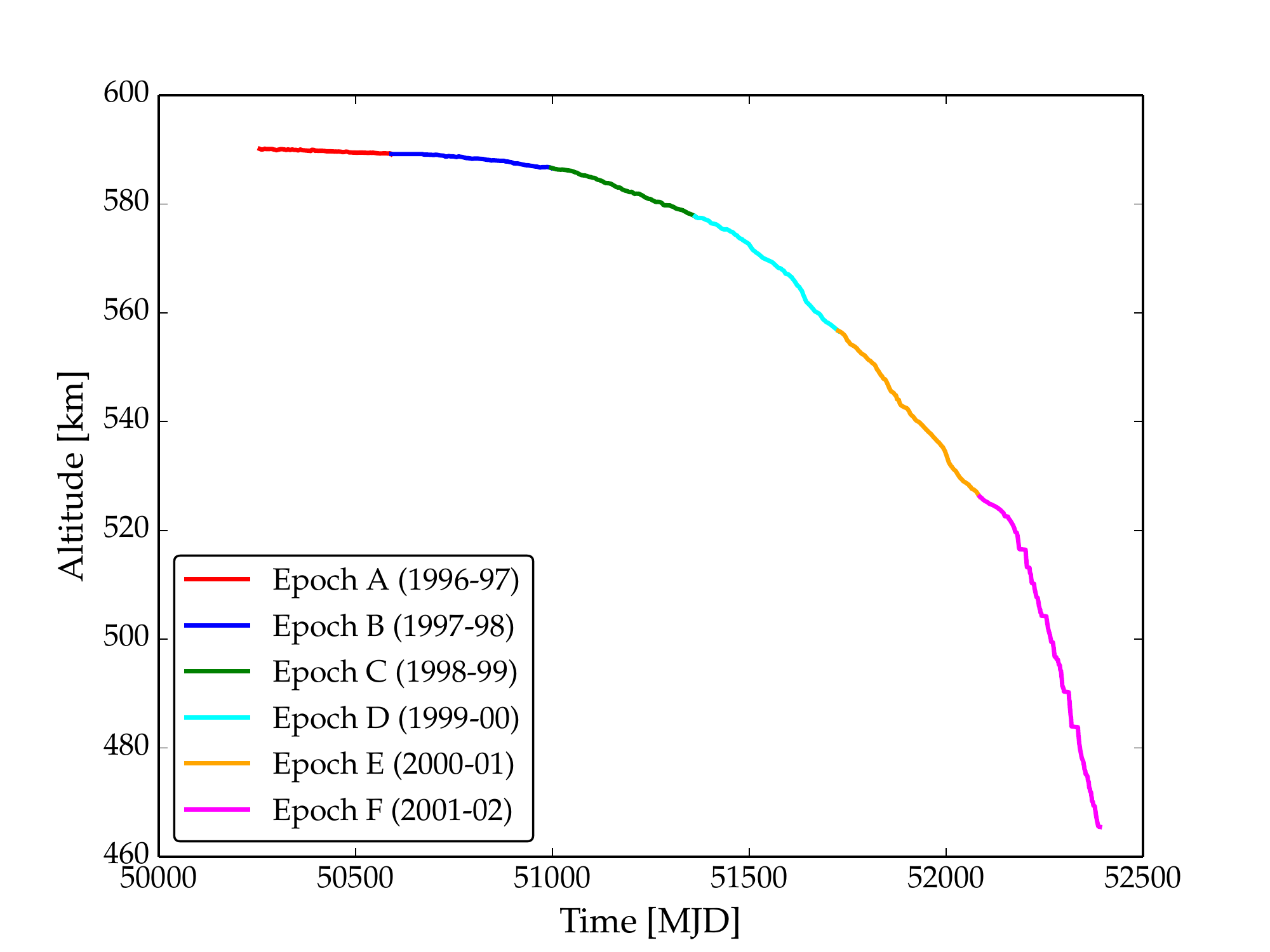}
\caption{\sax\ altitude vs. time, showing the satellite orbital decay. Different colours refers to the epochs defined in Table~\ref{table_epochs}.}
\label{fig_epoch_altitude}
\end{figure}

The Italian-Dutch \emph{Satellite per Astronomia X} \cite{boella97a} (Figure~\ref{fig_saxpm}), later renamed \sax\ in honour of Giuseppe ``Beppo'' Occhialini, was launched from Cape Canaveral on 1996 April 30. Its initial orbit was at an altitude of about 600~km with an inclination of 3.9$^{\circ}$. 
The satellite altitude decreased with time, due to the atmospheric drag (Figure~\ref{fig_epoch_altitude}) until \sax\ reentered the atmosphere seven years later, on 2003 April 29. The scientific payload, however, was switched off one year before, on 2002 April 30.

\sax\ carried onboard four Narrow Field Instruments: the Low and Medium Energy Concentrator System (LECS and MECS, \cite{parmar97,boella97b}) the High Pressure Gas Scintillator Proportional Counter (HPGSPC, \cite{manzo97}), and the Pho\-swich Detection System (PDS, \cite{frontera97}); besides the Gamma Ray Burst Monitor (GRBM, \cite{frontera97,costa98}) and the two Wide Field Cameras (WFC, \cite{jager97}).

During its lifetime, thanks to its large spectral coverage (0.1--300~keV), \sax\ contributed substantially to the advancement of X-ray astronomy, observing several classes of high-energy sources such as X-ray binaries, and discovering the afterglow of Gamma Ray Bursts (GRB) \cite{costa98,frontera03}. 

The Particle Monitor (PM, \cite{frontera97}, Figure~\ref{fig_saxpm}) was a part of the PDS experiment and was located inside the satellite, near the HPGSPC instrument. It consisted of a 2~cm diameter, 5~mm thick cylindrical plastic scintillator (BC-434) read-out by a photomultiplier tube (Hamamatsu R-1840). The detector was encapsulated in a 2~mm thick aluminium frame. The nominal energy threshold was about 1.2~MeV for electrons and 20~MeV for protons.
Since the photomultiplier electronic threshold was set below the signal amplitude corresponding to the minimum detectable particle energy, this setup ensures that the energy-integrated count rate is robust with respect to any possible variation in both the electronic threshold or gain.
The PM was used to switch off the high voltage supplies of the PDS instrument when the charged particle count rate, integrated in a 32~s time bin, exceeded a programmable threshold.
 
\begin{figure}[htbp]
\centering
\includegraphics[width=0.9\textwidth]{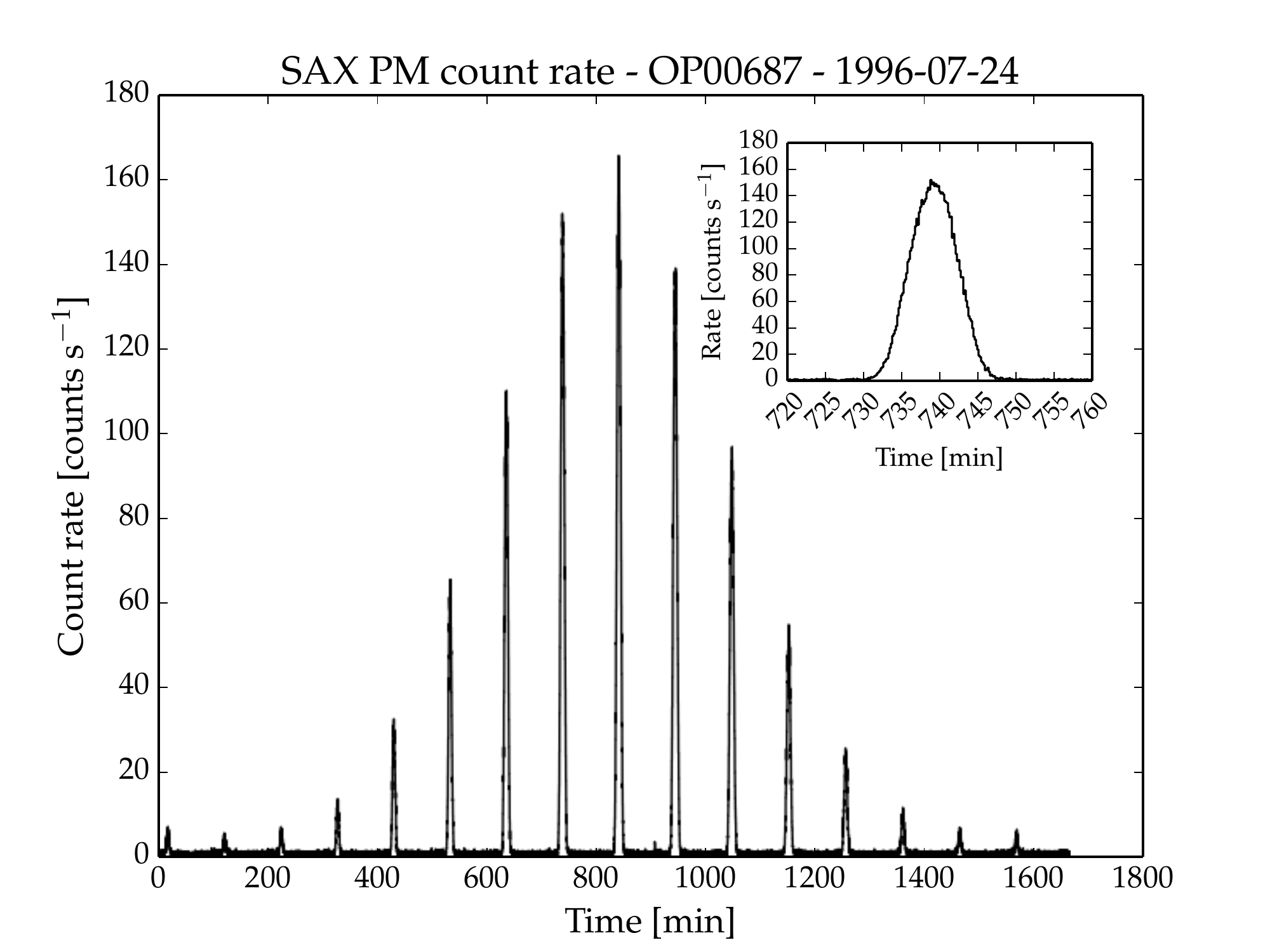}
\caption{\sax/PM count rate for the observation period OP00687, corresponding to July 24, 2006. The peaks in the count rate correspond to the various SAA passages. The inset shows in more detail one such passage.}
\label{fig_saa_rate}
\end{figure}

The Particle Monitor therefore provided continuous particle count rates along all the \sax\ lifetime. 
An example is shown in Figure~\ref{fig_saa_rate}, where the count rate in 10~s bins is shown during a \sax\ observation (Observation Period OP00687, starting July 24, 1996), spanning 1.2 days.
The average background level outside the SAA is about 2 counts s$^{-1}$, while in the SAA passages the count rate increases to more than a hundred counts per second.
The varying amplitude of the SAA flux is due to the satellite orbital plane drifting with respect to the Earth surface, resulting in different regions sampled at different times, e.g. during successive orbits.

\section{The SAX Particle Monitor data}\label{pm}

All the data produced by the instruments onboard \sax\ were inserted in so-called Source Packets by the On Board Data Handling (OBDH) subsystem. 
These Source Packets were then transmitted to ground through eight Virtual Channels (VC).
VC0 transmitted frames containing On Board Time (OBT) conversion
information, that was subsequently processed by the Operational Control
Center (OCC) to convert time in UTC.
The \emph{housekeeping} Source Packets
(\texttt{PDHK000} packet type for the PDS experiment) were transmitted on VC1:
among them, the spacecraft attitude data (processed by the OCC for the
orbital reconstruction), and the PM count rates.
These data were sampled every 1~s.
The VC from 2 to 7 dealt with Source Packets produced by the scientific
instruments.  
However, in our analysis we used the PM rates transmitted through VC1 (the so-called \texttt{PMEV}),
because during the SAA passages the scientific instruments were switched
off, while data from the VC1 were always recorded.
In order to increase the signal to noise ratio, the PM data were rebinned at
10~s.  

\subsection{SAA passage duration}\label{pmpass}
The first step in the analysis consists in the subdivision of the \sax/PM dataset in 6 epoch intervals, each corresponding to roughly one year, reported in Table~\ref{table_epochs}.
There is a data gap of about two months around 1997 July, when the satellite went into safe mode due to a gyroscope failure.

\begin{table}[htdp]
\caption{BeppoSAX observation epochs used in the analysis.}
\begin{center}
\begin{tabular}{ccccc}
\hline
Epoch & OP interval & Start & End & Altitude range (km)\\ \hline
A & 00500--02025 & 05/1996 & 06/1997 & 583.8--594.1\\
B & 02026--04853 & 07/1997 & 06/1998 & 584.6--593.2\\
C & 04854--07129 & 07/1998 & 06/1999 & 577.0--588.7\\
D & 07130--09353 & 07/1999 & 06/2000 & 556.7--578.4\\
E & 09354--11425 & 07/2000 & 06/2001 & 526.2--557.1\\
F & 11426--13000 & 07/2001 & 04/2002 & 464.3--528.3\\ 
\hline
\end{tabular}
\end{center}
\label{table_epochs}
\end{table}

From the PM count rate time series, we extracted the duration of SAA passages, defined as the time length in which the count rate rises above a given threshold (equivalent, in this case, to the average background level) for more than 50 seconds.

\begin{figure}[htbp]
\centering
\includegraphics[width=0.9\textwidth]{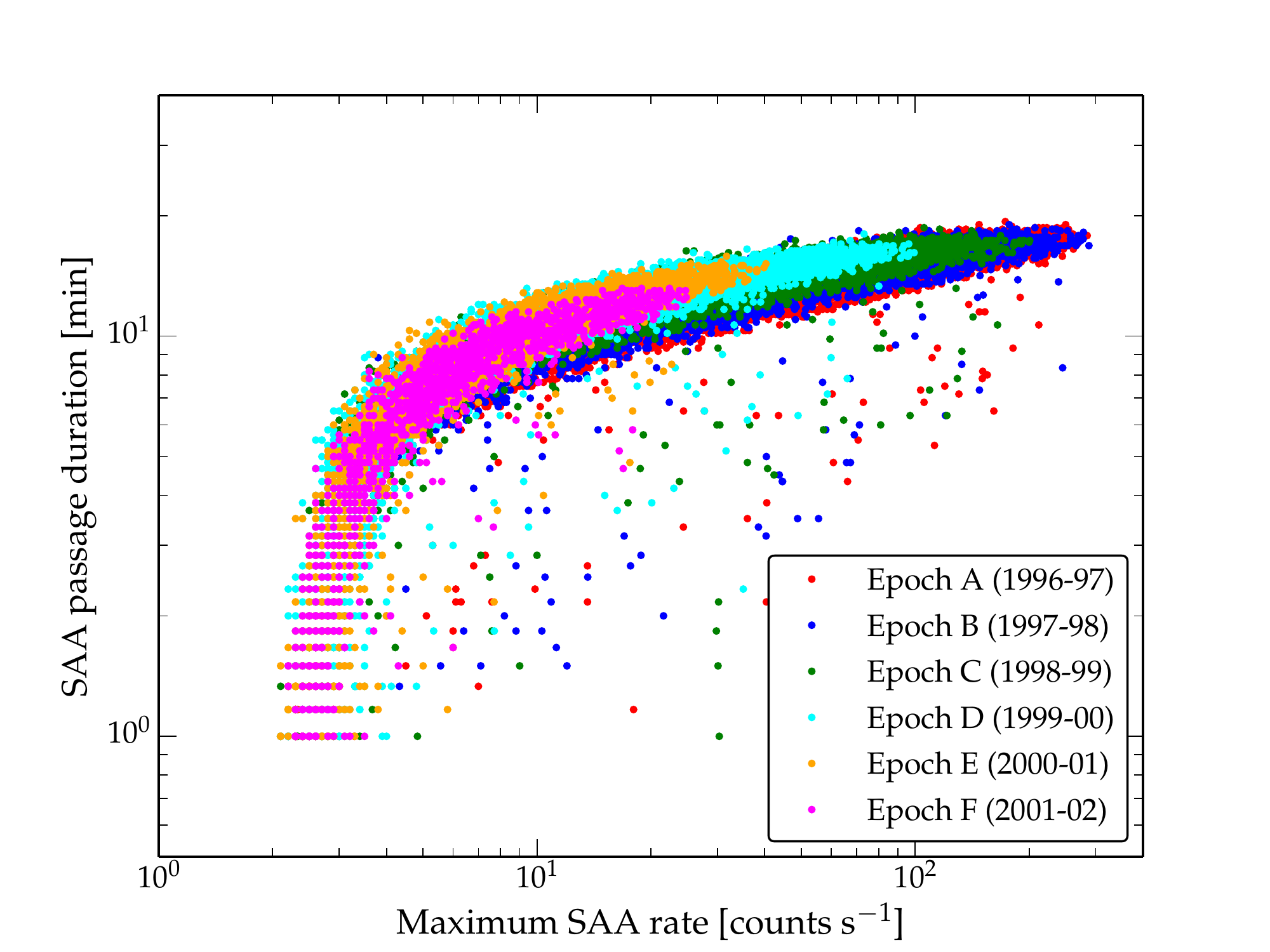}
\caption{SAA passage duration versus peak \sax/PM count rate in the passage for the various epochs.}
\label{fig_saa_dvr}
\end{figure}

In Figure~\ref{fig_saa_dvr} the duration of each SAA passage is shown for the whole \sax\ dataset, compared with the peak count rate in the passage. The duration depends on the depth, and varies from 1 to 20 minutes. Since shorter passages usually correspond to the grazing of SAA edges, the maximum particle flux is correspondingly lower.
%

\begin{figure}[htbp]
\centering
\includegraphics[width=0.9\textwidth]{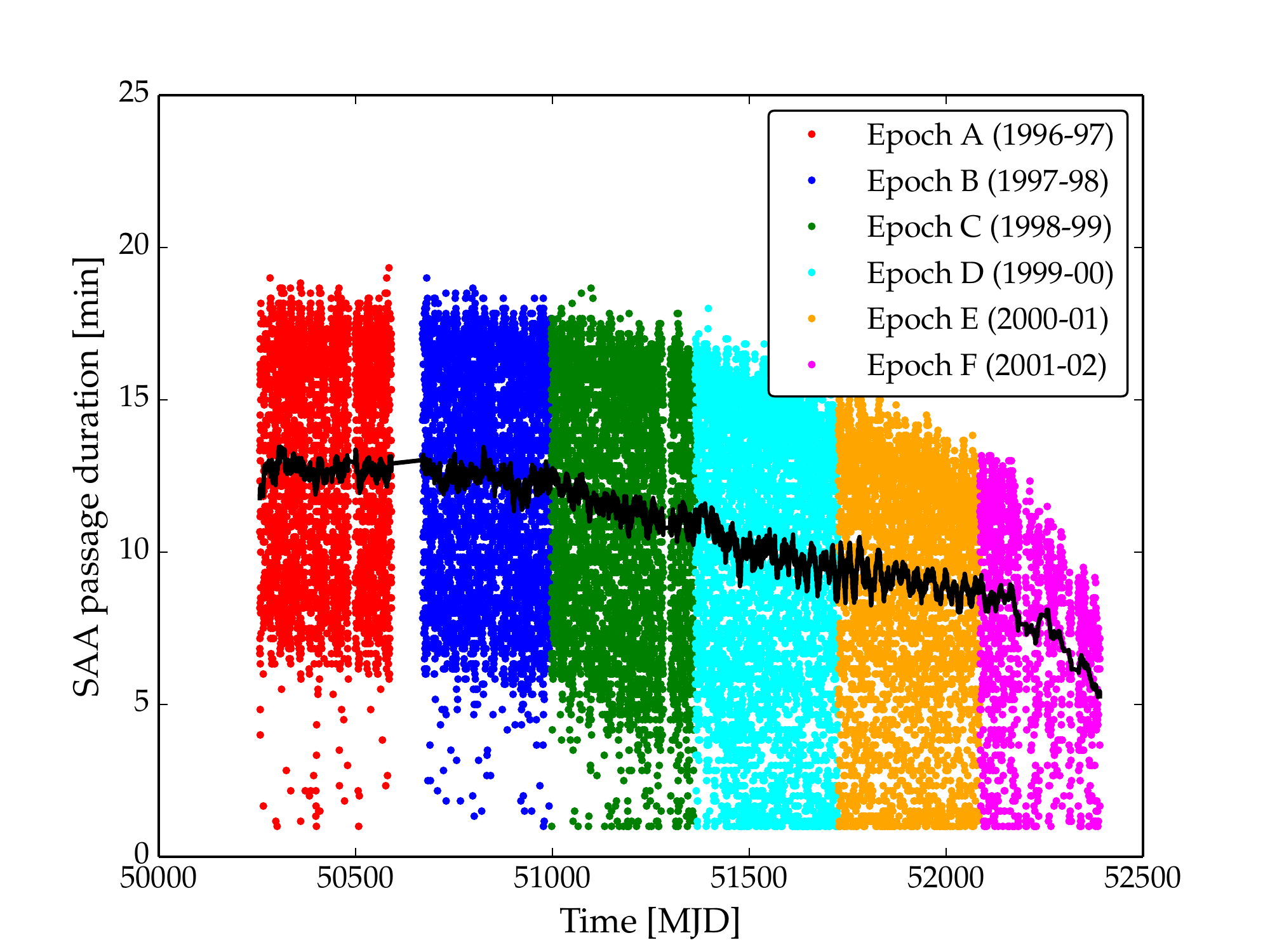}
\caption{SAA passage duration versus time for the various epochs. The thick black line is a moving average over 100 adjacent points (i.e. successive passages).}
\label{fig_saa_dvt}
\end{figure}

\begin{figure}[htbp]
\centering
\includegraphics[width=0.9\textwidth]{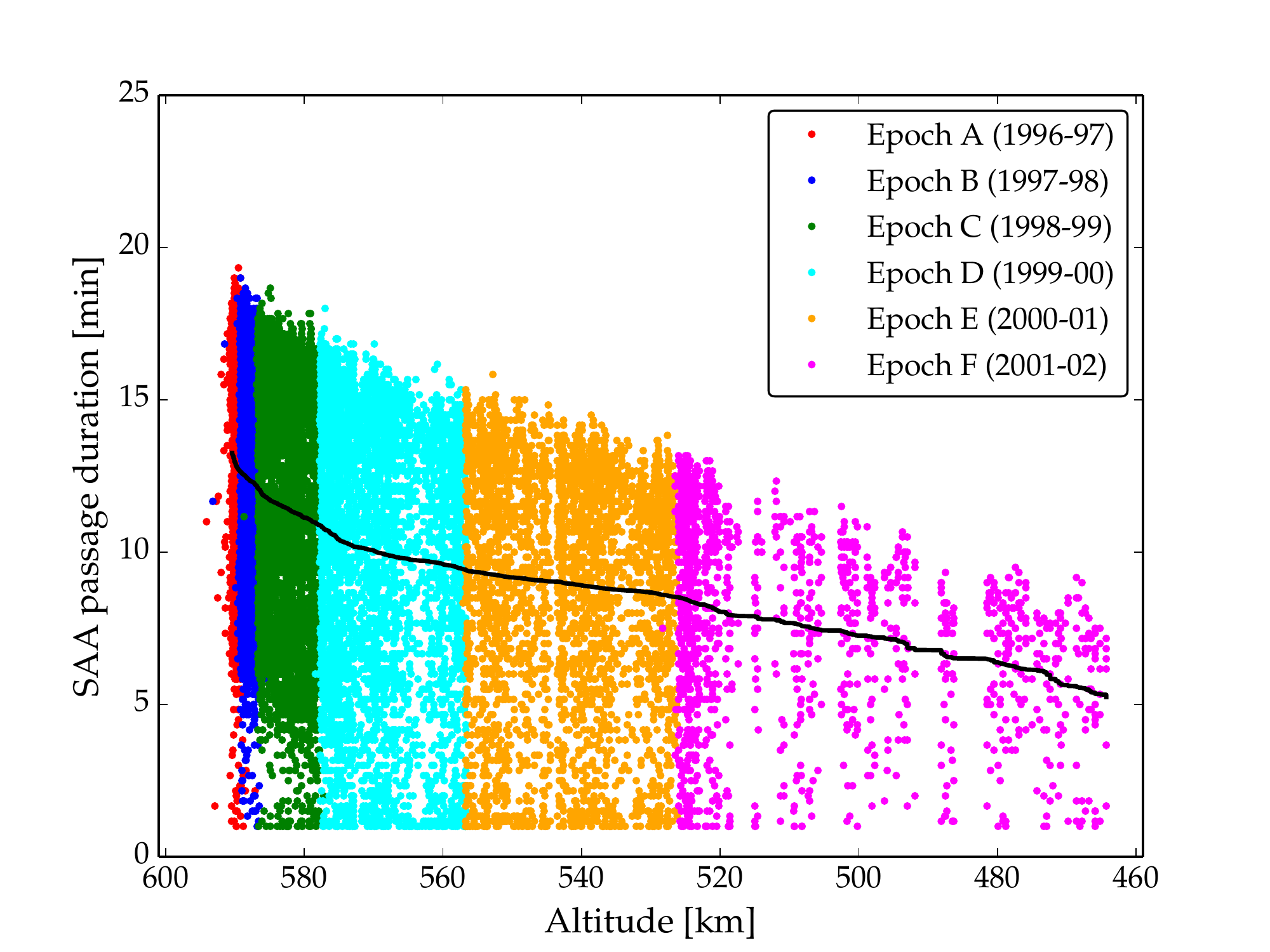}
\caption{SAA passage duration versus mean OP altitude for the various epochs. The thick black line is a moving average over 100 adjacent points.}
\label{fig_saa_dvh}
\end{figure}

\begin{figure}[htbp]
\centering
\includegraphics[width=0.9\textwidth]{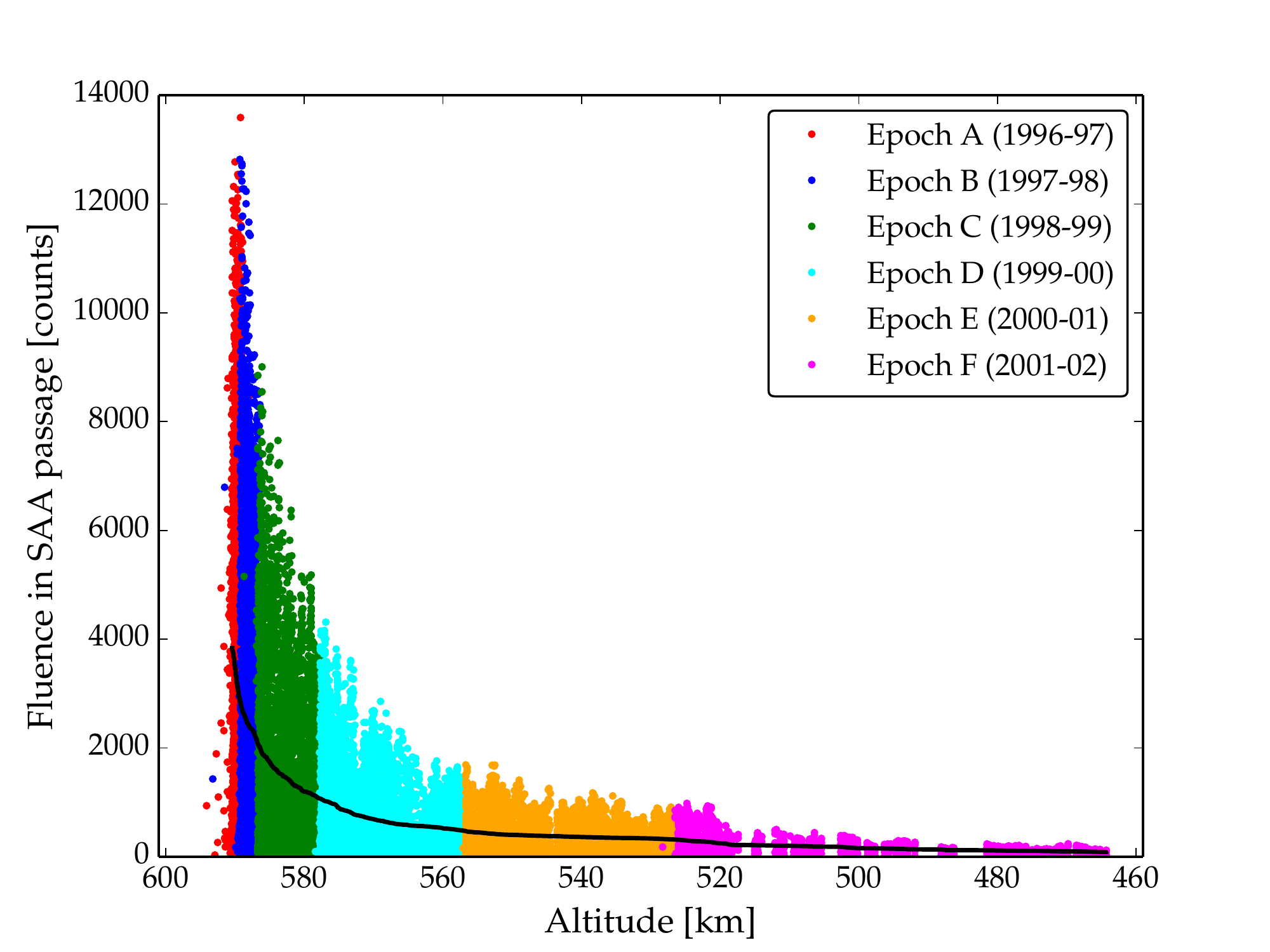}
\includegraphics[width=0.9\textwidth]{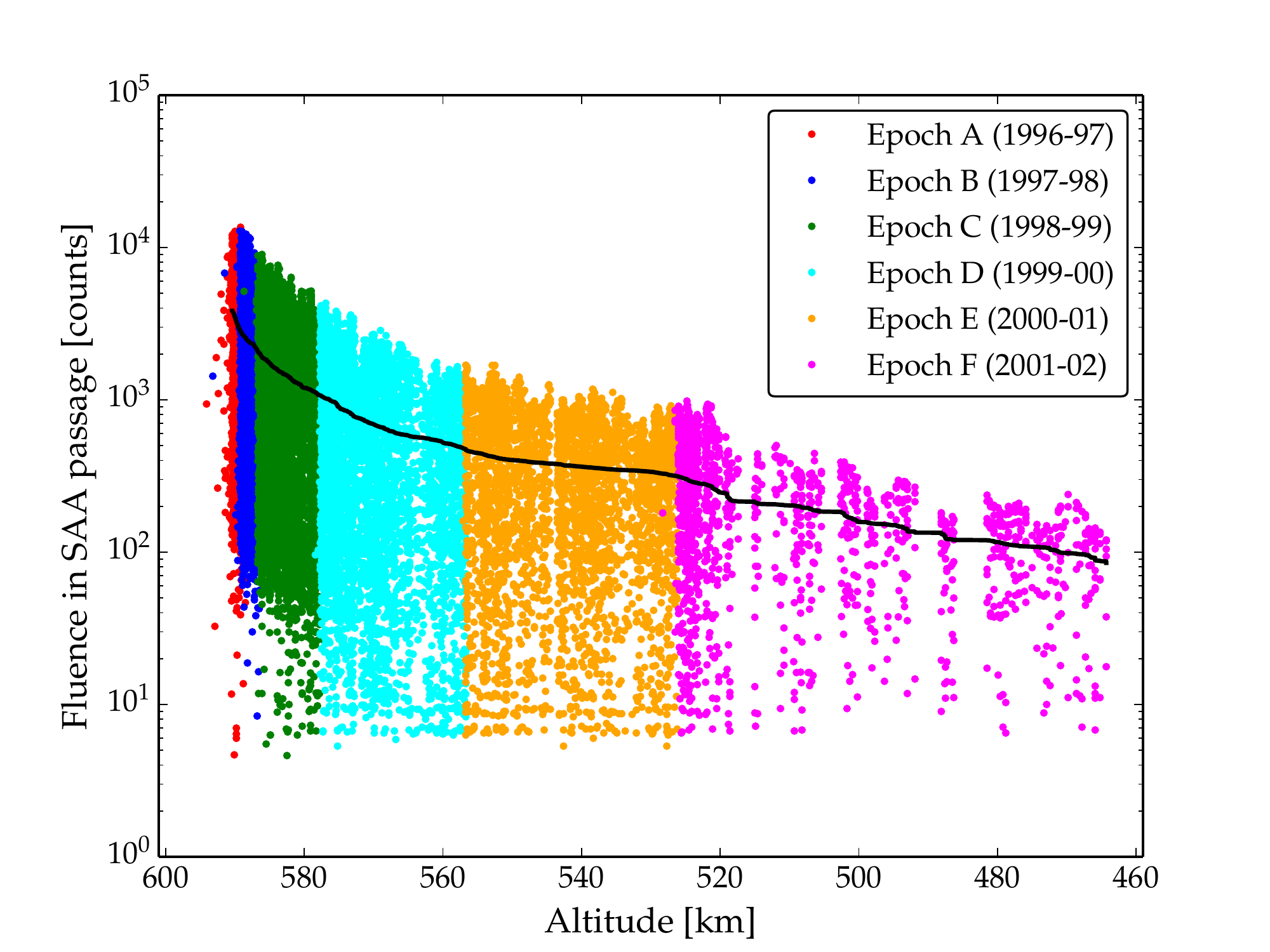}
\caption{SAA passage fluence measured by \sax/PM versus mean OP altitude for the various epochs, in linear (\emph{upper panel}) and log scales (\emph{lower panel}). The thick black line is a moving average over 100 adjacent points.}
\label{fig_saa_fvh}
\end{figure}

In Figure~\ref{fig_saa_dvt} the SAA passage durations are plotted as a function of their epoch. The average passage duration decreases with time. This is due to the lower satellite altitude as shown in Figure~\ref{fig_saa_dvh}.
An interesting quantity is the \emph{fluence} of high-energy particles in the SAA, i.e. the total counts accumulated during a passage. This quantity relates directly to the orbital radiation damage of a high-energy detector. In Figure~\ref{fig_saa_fvh} the fluence for each SAA passage is shown as a function of the altitude: lowering the altitude from $\sim$590 to $\sim$550 km leads to a dramatic fluence decrease by a factor of $\sim$10. 

\subsection{SAA count rate maps}\label{pmmaps}

By plotting the PM data along a latitude-longitude grid it is possible to derive count rate maps that show the SAA extent at various epochs/altitudes.
In order to have a sufficient signal to noise ratio, spatial bins of 1$^{\circ}$ in latitude and longitude were chosen.

\begin{figure}[htbp]
\centering
\includegraphics[width=0.7\textwidth]{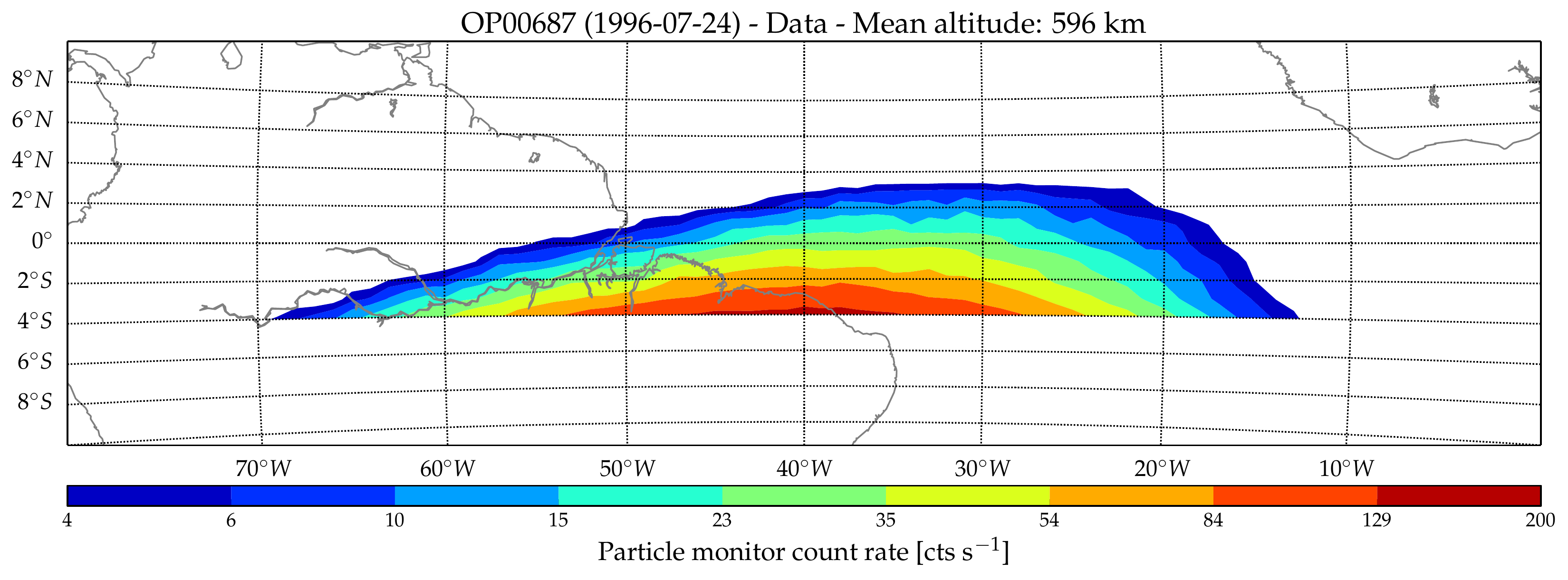}
\includegraphics[width=0.7\textwidth]{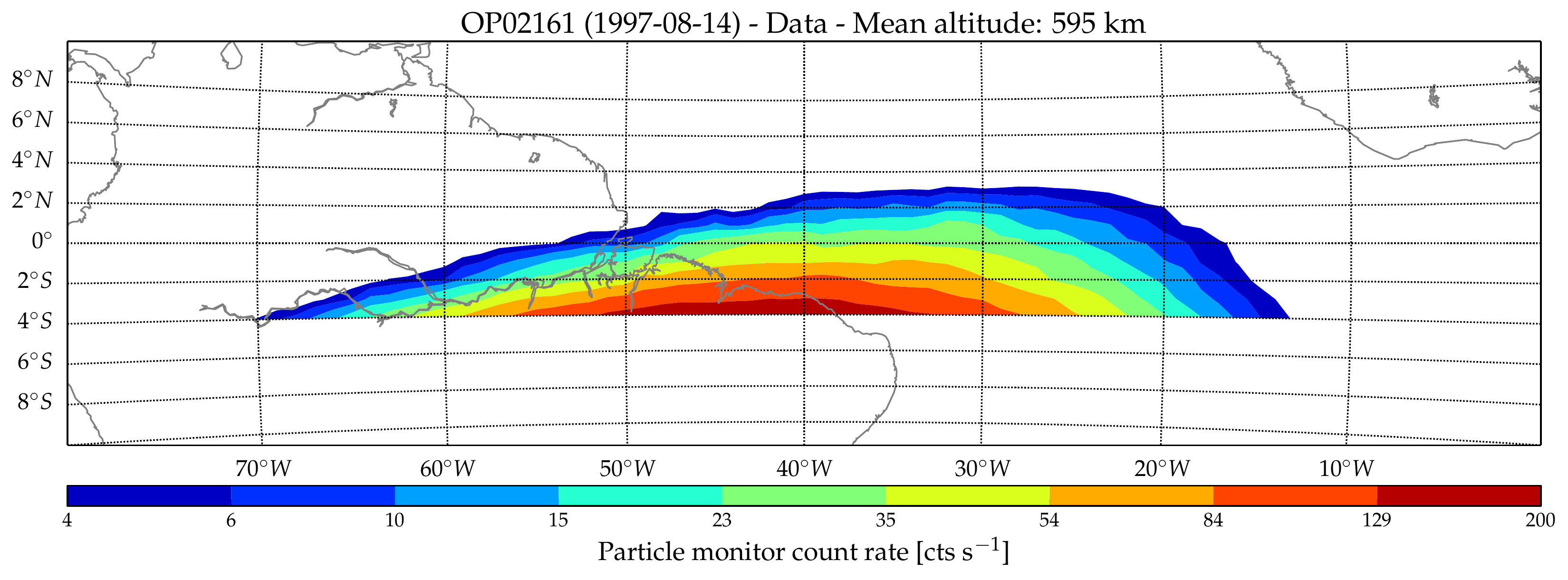}
\includegraphics[width=0.7\textwidth]{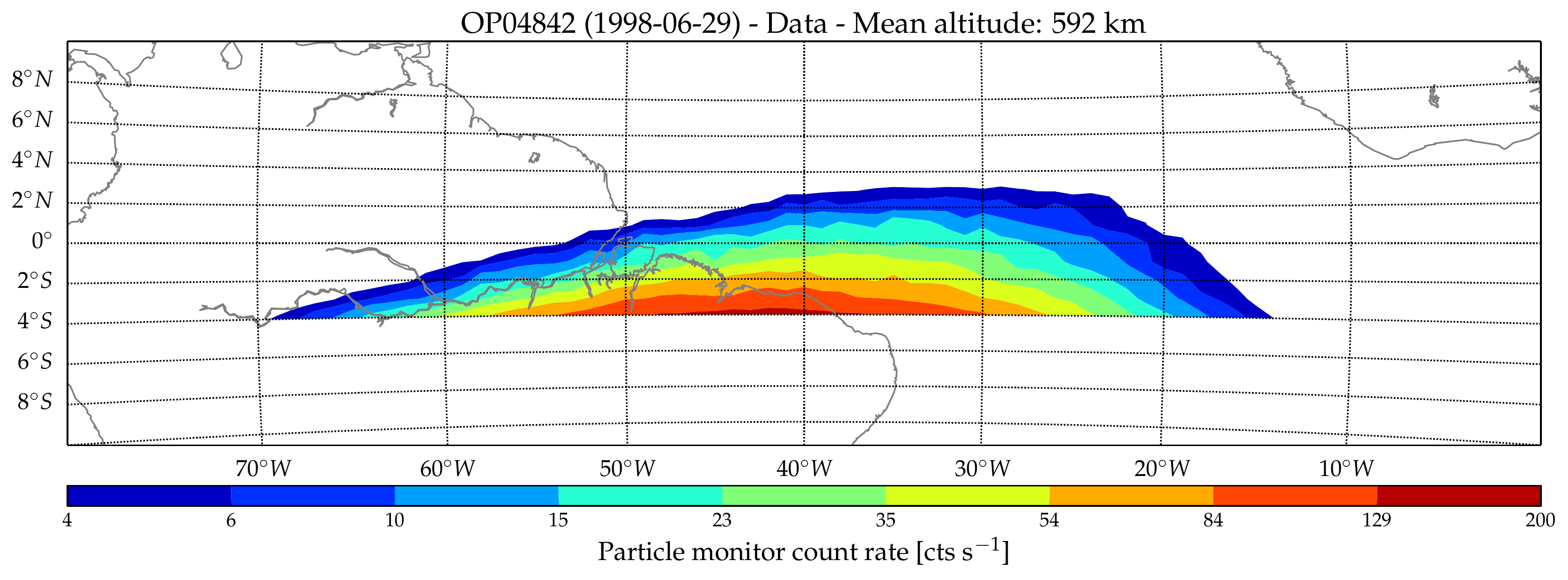}
\includegraphics[width=0.7\textwidth]{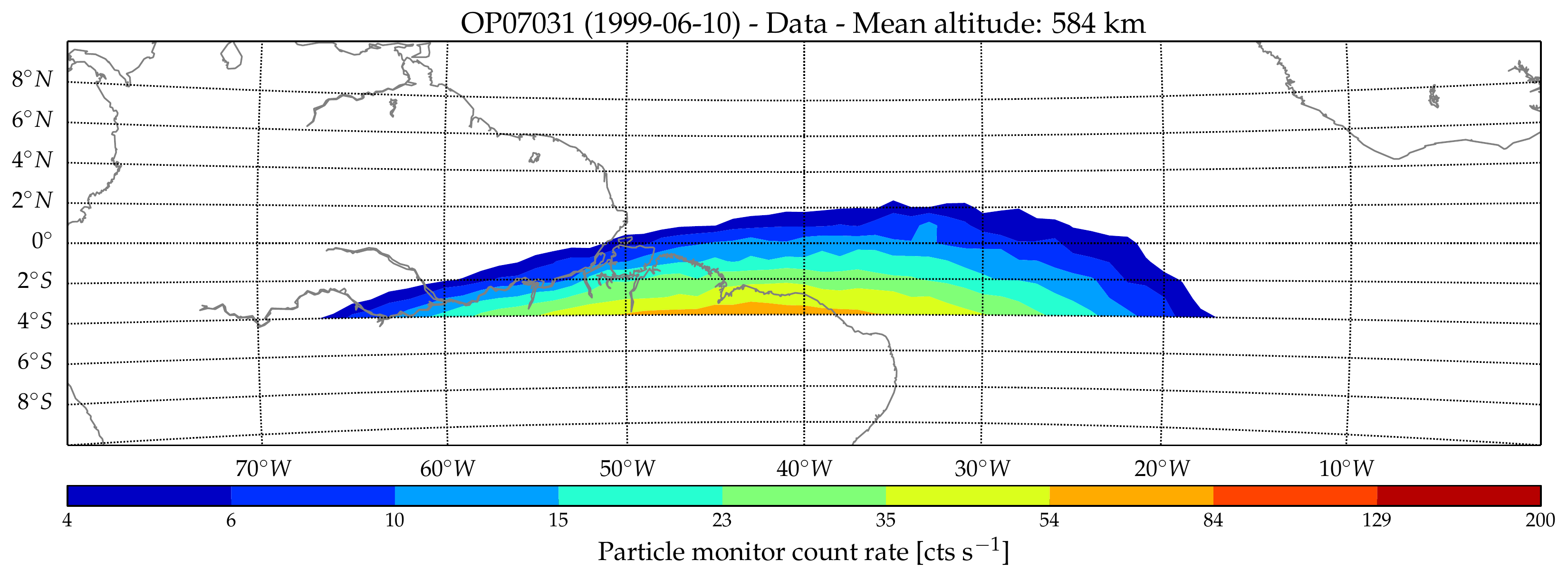}
\includegraphics[width=0.7\textwidth]{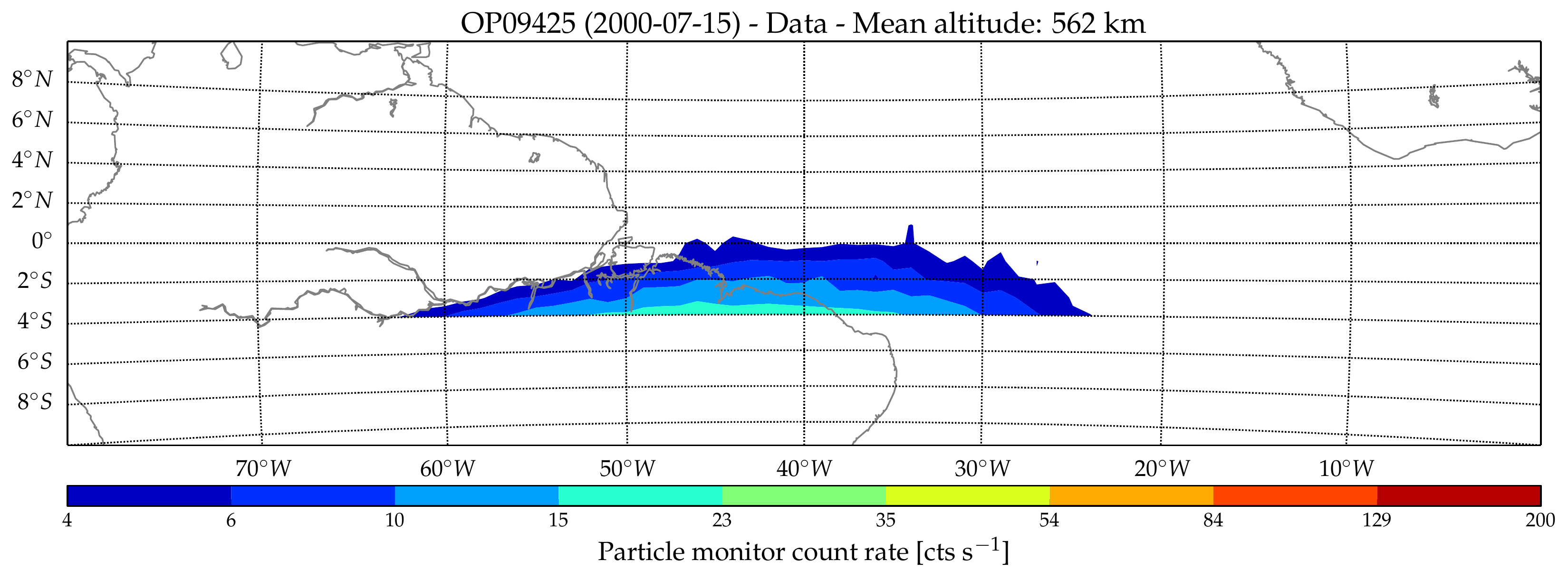}
\includegraphics[width=0.7\textwidth]{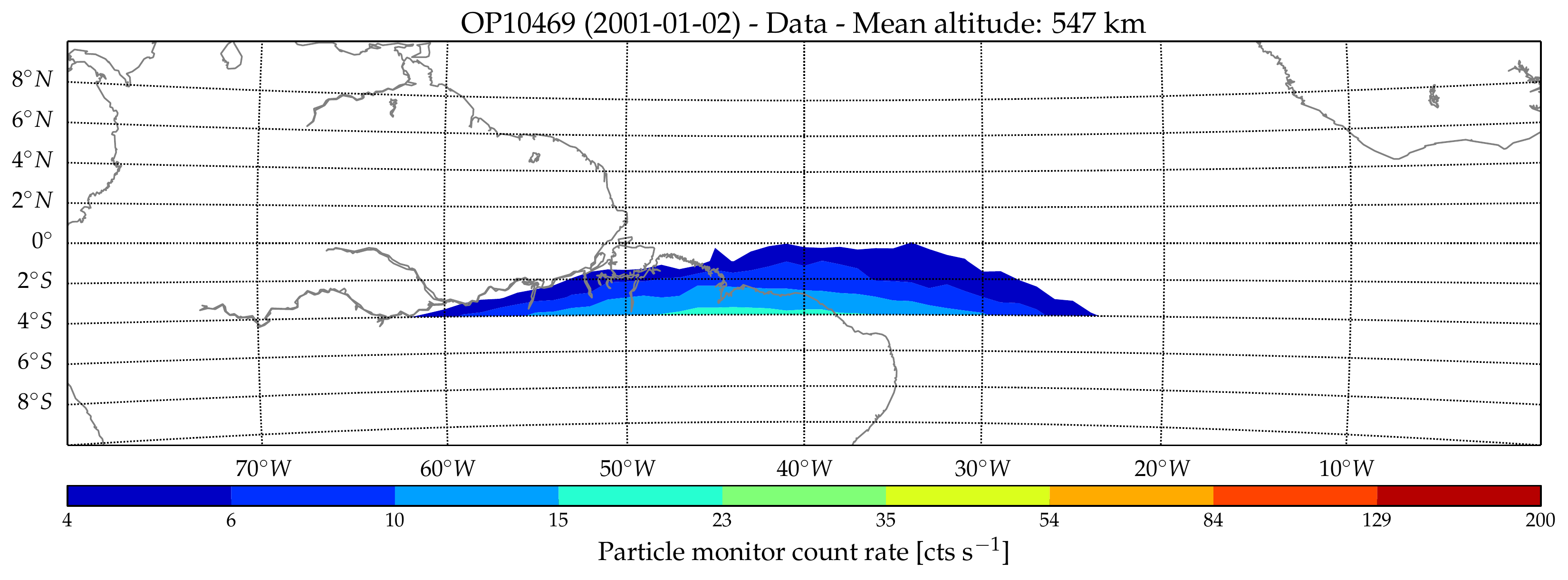}
\includegraphics[width=0.7\textwidth]{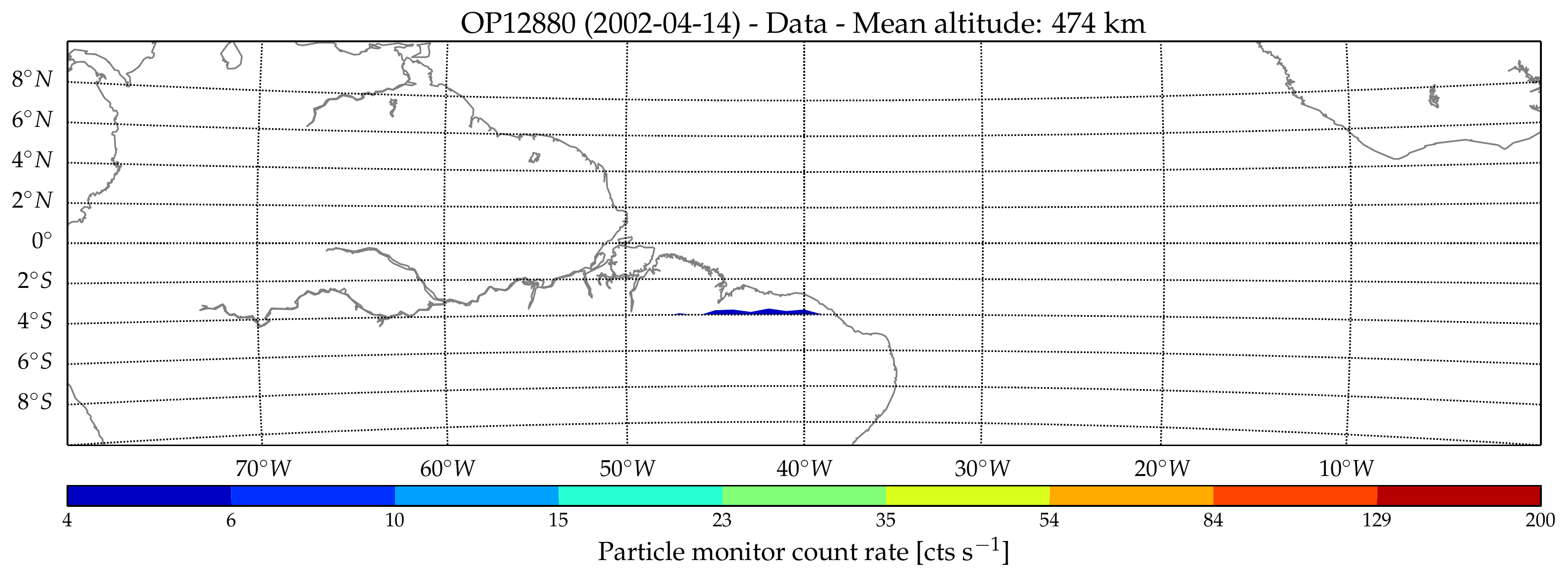}
\caption{SAA maps for various representative epochs. From above to below, data from 1996 to 2002 and from a mean altitude of $\sim$600 to $\sim$470 km.}
\label{fig_saa_pm_maps}
\end{figure}

The dramatic decrease in fluence shown in Figure~\ref{fig_saa_fvh} is also well evident in the panels of  Figure~\ref{fig_saa_pm_maps}: both the SAA width and peak fluxes decrease with altitude. Below $\sim$520~km, only the lowest latitude bins show a significant flux above the background.

In order to quantitatively estimate this decrease, the count rate maps can be analyzed in a similar way as F\"urst et al. \cite{furst09} have done for RXTE data, i.e. using a Weibull function to describe the longitude-dependent profile:
\begin{equation}\label{eq_weibull}
W(x) = \frac{Ak}{\lambda} \left(\frac{x - \theta}{\lambda} \right)^{k-1}
   \exp\left[-\left(\frac{x - \theta}{\lambda} \right)^k\right]
   \mbox{\, for $x\ge \theta$}
\end{equation}
\begin{equation}\label{eq_weibullbis}
W(x) = 0  \mbox{\, for $x< \theta$}
\end{equation}
where $x$ is the longitude, $A$ is a normalization factor, $\lambda$ represents a characteristic scale length, $k$ governs the shape and  $\theta$ controls the position of the maximum $x_M$ of the function:
\begin{equation}\label{eq_weibull2}
x_M = \lambda \left(\frac{k-1}{k}\right)^{\frac{1}{k}} + \theta
\end{equation}
The skewed Weibull function \cite{brown95,lai14}, usually employed to describe many types of physical phenomena (e.g. the distribution in size of particles resulting from the crushing of materials), gives a good phenomenological description of the asymmetry of the longitude profile of the SAA, and allows to describe its overall shape with a few parameters.

\begin{figure}[htbp]
\centering
\includegraphics[width=0.9\textwidth]{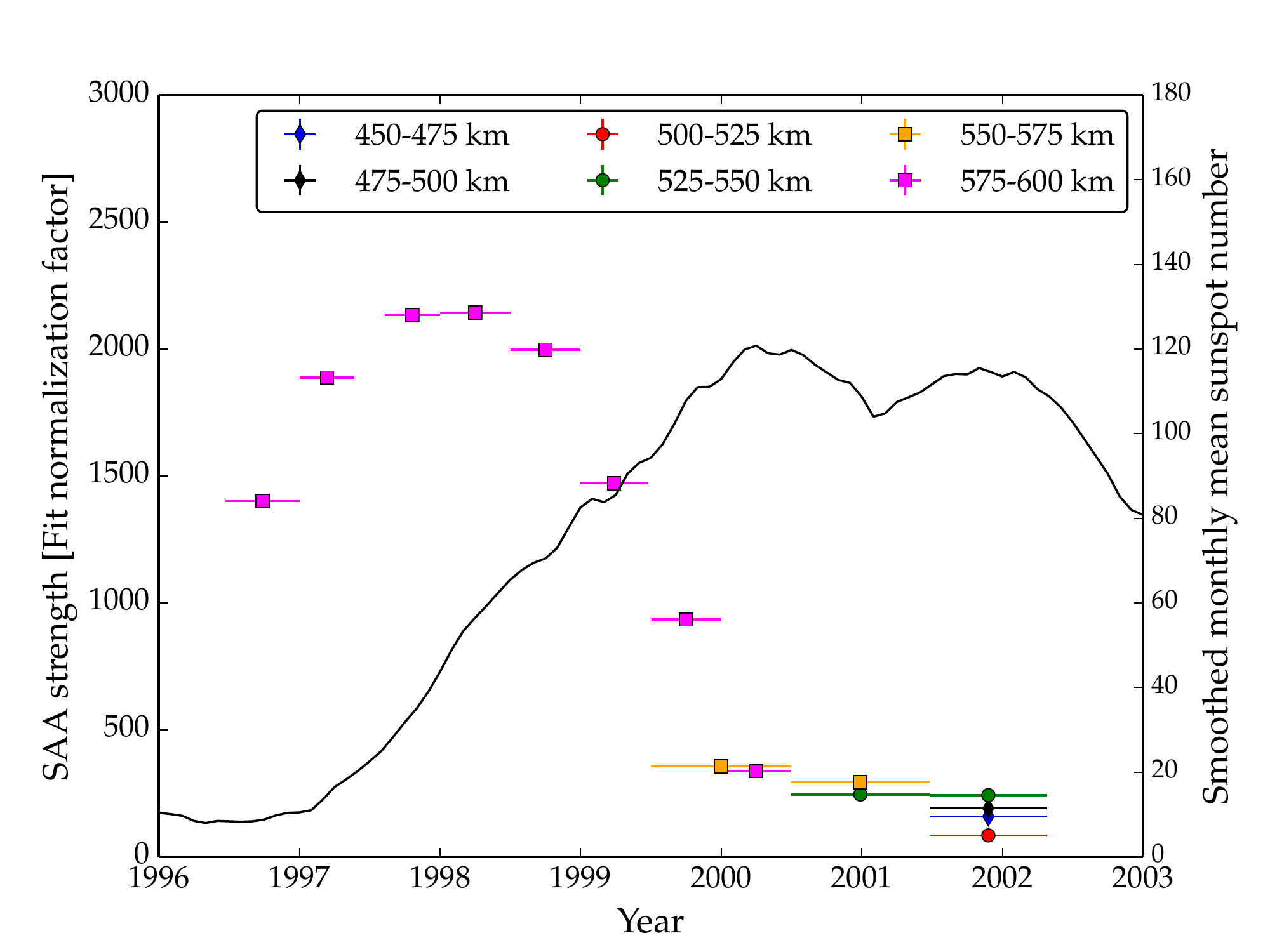}
\caption{The SAA strength, proportional to the normalization factor $A$ of Eq.~\ref{eq_weibull}. The solid black line is the average monthly sunspot number, a proxy of the solar activity.}
\label{fig_saa_strength}
\end{figure}

\begin{figure}[htbp]
\centering
\includegraphics[width=0.9\textwidth]{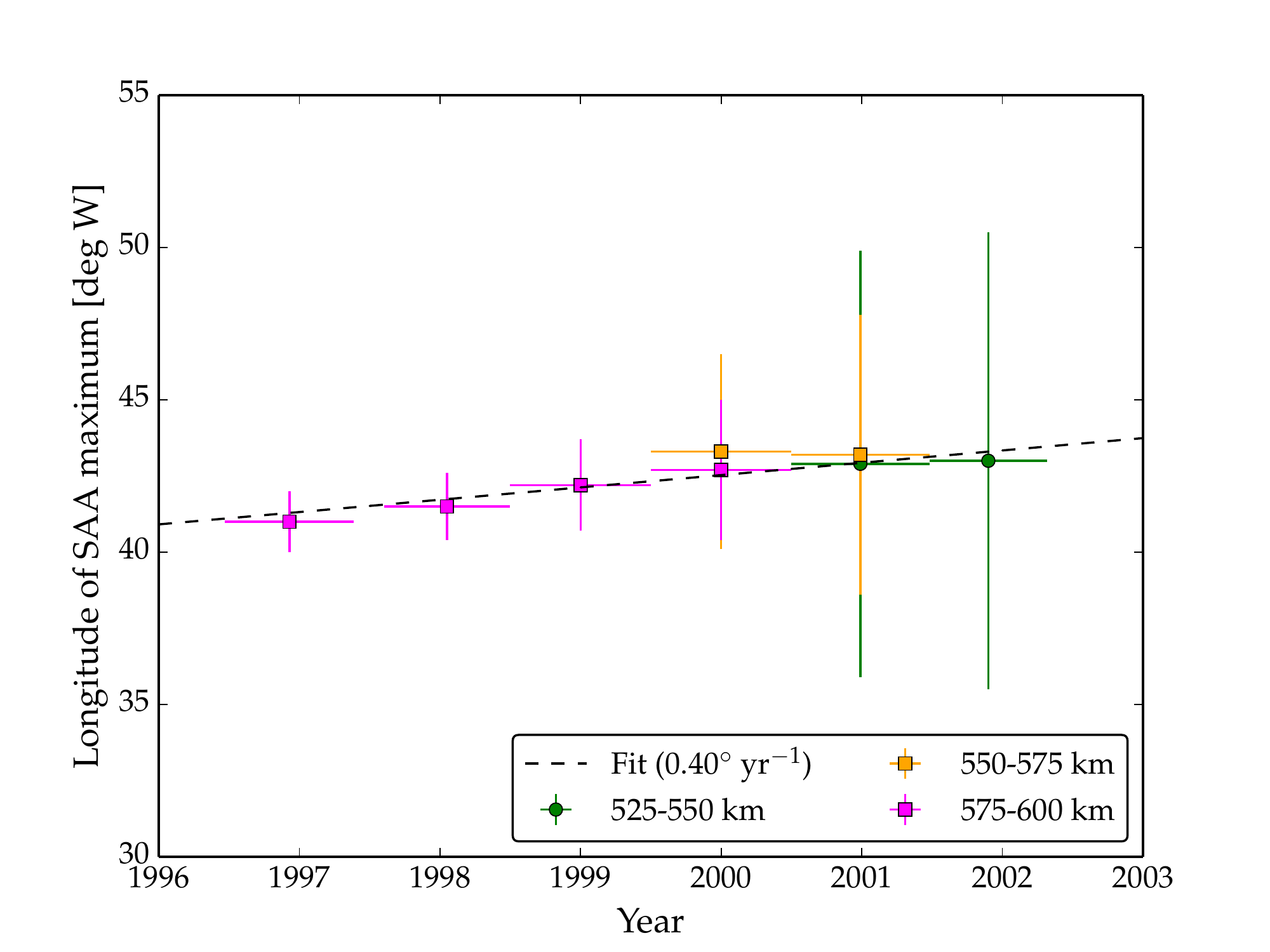}
\caption{The location of the SAA maximum determined from fitting the count rate maps with Eq.~\ref{eq_weibull}. The westward drift is well apparent. The dashed line is the linear fit, corresponding to 0.40$^{\circ}$~yr$^{-1}$.}
\label{fig_saa_westward}
\end{figure}

The longitudinal profile in correspondence of the southernmost latitude bin ($-4^{\circ}$) has been fitted using Eq.~\ref{eq_weibull}. The results are shown in Figures~\ref{fig_saa_strength} and \ref{fig_saa_westward}, that report the fitted $A$ factor and the $x_M$ values, as a function of the epoch and of the altitude (in 25~km intervals). 
The strength decreases with epoch and altitude, confirming the trend shown in Figure~\ref{fig_saa_fvh}.
The maximum flux around the first months of 1998, seen also by RXTE \cite{furst09}, is in anti-correlation with the Solar activity, quantified e.g. by the 10.7 cm radio flux from the Sun or, as shown in Figure~\ref{fig_saa_strength}, the smoothed average number of sunspots\footnote{Data from \url{http://www.ngdc.noaa.gov/stp/space-weather/solar-data}.}.
The delay between the solar cycle minimum and the maximum SAA strength is of $\sim$1 year. 
In Figure~\ref{fig_saa_strength}, in order to better show the peak for the normalization factor, for the highest altitude bin (575--600~km) epochs A, B and C were splitted into two subperiods ($\sim$6 month long) each.
Higher solar activity heats the upper atmosphere, thus lowering the trapped particle flux by absorption and deflection \cite{harris62,dachev99}.

There is a good correlation between \sax/PM and RXTE data, acquired over the same epochs.
The westward drift of the SAA maximum is $(0.40 \pm 0.08)^{\circ}$~yr$^{-1}$, compatible with the 0.346$^{\circ}$~yr$^{-1}$ drift seen at the much lower latitude of $-23^{\circ}$ by RXTE in 1998-2003.

\subsection{Local magnetic field}\label{pmmagfield}
\begin{figure}[htbp]
\centering
\includegraphics[width=0.9\textwidth]{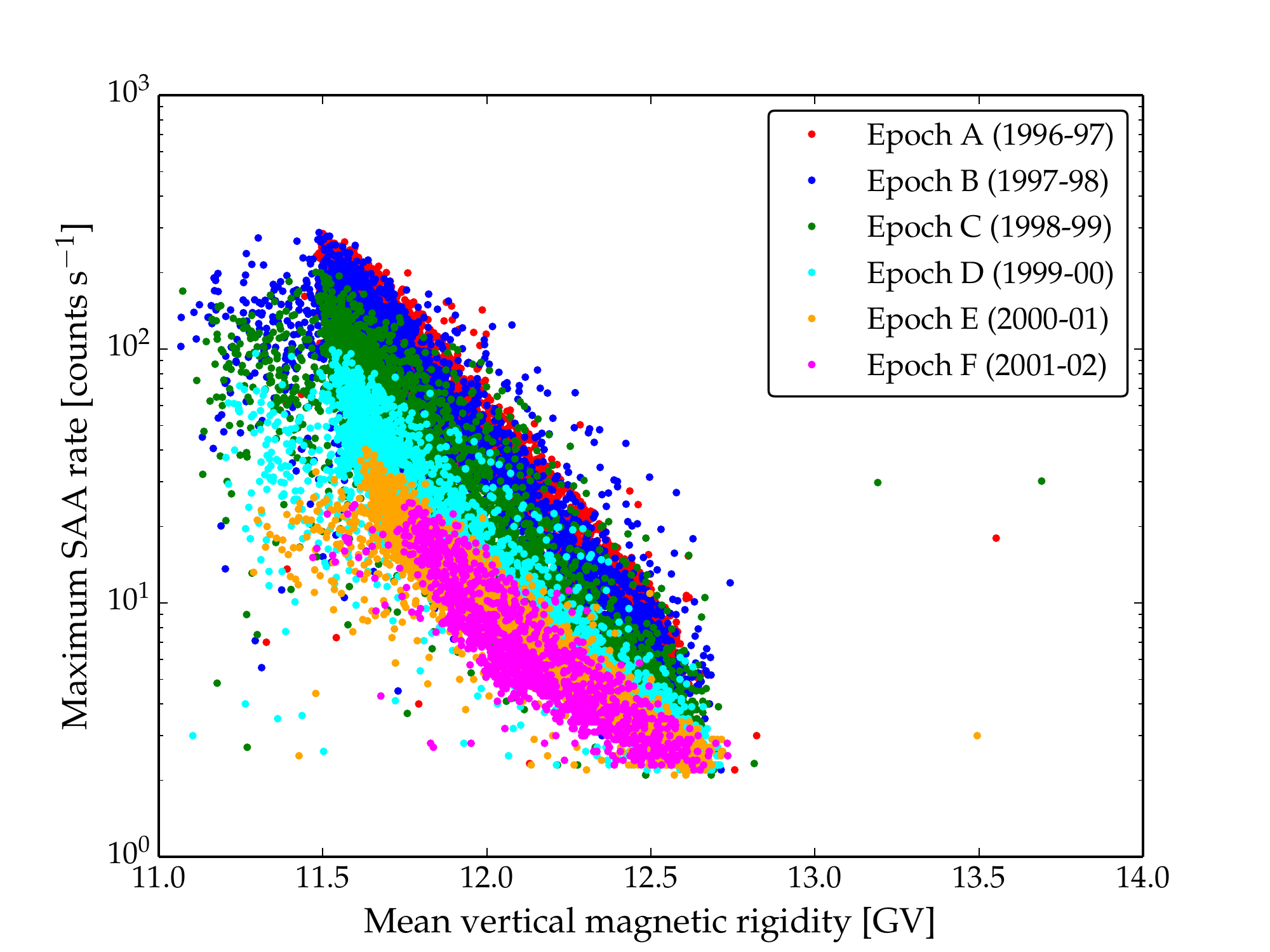}
\caption{Correlation between the mean vertical magnetic rigidity in a SAA passage and the maximum count rate observed by \sax/PM.}
\label{fig_saa_rigidity}
\end{figure}

Another interesting comparison can be made with the local magnetic field properties.
For each data point, the local values of the geomagnetic field have been calculated from the International Geomagnetic Reference Field (IGRF11, \cite{finlay10}) model.
As an example, in Figure~\ref{fig_saa_rigidity} the maximum count rate observed by \sax/PM in a SAA passage is seen to anti-correlate with the mean vertical magnetic rigidity, calculated from the St\"ormer formula
\begin{equation}\label{stormer}
R_c = 14.5 \times \left( 1 + \frac{h}{R_E} \right)^{-2} \cos^4\theta_M \mbox{\,\, GV}
\end{equation}
where $\theta_M$ is the geomagnetic latitude, $h$ is the altitude and $R_E$ is the mean Earth radius. 
Cosmic-ray particles having a rigidity lower than the one given in the previous equation cannot reach this location in the magnetosphere.
Subsequent epochs in Figure~\ref{fig_saa_rigidity} show similar power-law behavior in the rigidity vs. rate plane. The normalisation decreases with the altitude. 

\begin{figure}[htbp]
\centering
\includegraphics[width=0.9\textwidth]{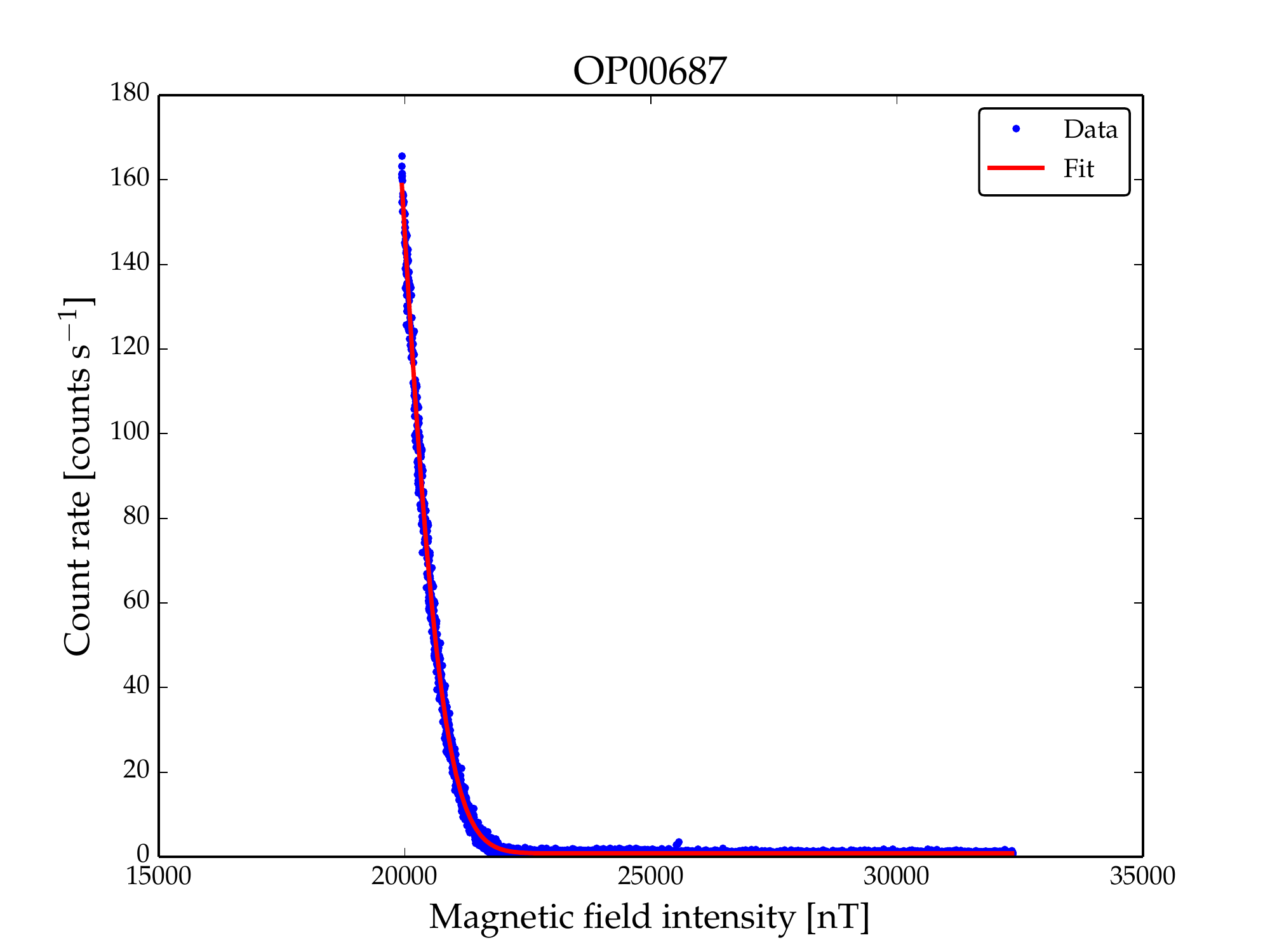}
\caption{Local magnetic field intensity versus \sax/PM count rate, for an example observation period (OP00687, July 24, 2006, the same as Figure~\ref{fig_saa_rate}). The Gaussian fit with Eq.~\ref{gaussfit} is also shown.}
\label{OP00687_Bfield}
\end{figure}

The South Atlantic Anomaly is characterised by a lower local magnetic field intensity. This is well apparent in Figure~\ref{OP00687_Bfield}, where the count rate values for a representative OP (00687) are well fitted by the following Gaussian function:
\begin{equation}\label{gaussfit}
F_B = \frac{K}{\sqrt{2\pi}\sigma} \exp{-\frac{(B-B_0)^2}{2\sigma^2}}
\end{equation}
where $F_B$ is the \sax/PM count rate (in counts s$^{-1}$) and $B$ is the magnitude of the magnetic field, in nT (1 nT = 10$^{-5}$ G).
In other words, the lower the field intensity the higher is the flux of trapped particles. For this particular OP, the fit parameters are $B_0 = (18960 \pm 14)$ nT, $\sigma = (888 \pm 4)$ nT and $K = (64.7 \pm 1.1) \cdot 10^{4}$ counts s$^{-1}$ nT.

The parameters $B_0$, $\sigma$ and $K$ are slightly time-dependent. 
A good description, accurate to a few percent, is given by linear fit over all \sax/PM observations, i.e. using the relation:
\begin{equation}\label{e:linfit}
y = a(T-T_0) + b 
\end{equation}
where $y$ stands for $B_0$, $\sigma$ or $K$; $T$ is the OP average epoch (in MJD) and $T_0$ is MJD 50000. The parameter values of Eq.~\ref{e:linfit}, for each of the Gaussian function parameters, are given in the following Table~\ref{tab_fit}.

\begin{table}[htdp]
\caption{Long-term variation of the relation between \sax/PM count rate and the local magnetic parameters. The table reports the linear fit parameters $B_0 = a(T-T_0) + b$, $\sigma = a(T-T_0) + b$ and $K = a(T-T_0) + b$, respectively, where $T$ is the OP average epoch (in MJD) and $T_0$ is MJD 50000.}
\begin{center}
\begin{tabular}{ccc}
Parameter & $a$ & $b$ \\ \hline
$B_0$ & $(-0.30 \pm 0.03)$ nT yr$^{-1}$ & $(19500 \pm 1300)$ nT \\
$\sigma$ & $(0.07 \pm 0.01)$ nT yr$^{-1}$ & $(730 \pm 30)$ nT \\
$K$ & $(-190 \pm 20)$ cts s$^{-1}$ yr$^{-1}$ nT & $(46.3 \pm 0.1) \cdot 10^4$ cts s$^{-1}$ nT \\ \hline
\end{tabular}
\end{center}
\label{tab_fit}
\end{table}

\section{Conclusions}\label{conclusions}
We have shown the shape and behaviour of the South Atlantic Anomaly as sampled by the high-energy particle flux measured by the Particle Monitor onboard \sax.
Its flux and extent are strongly dependent on altitude, decreasing by about one order of magnitude from 600~km to 550~km, on the phase of the solar cycle and on local magnetic field properties as intensity and mean cutoff rigidity. The location of the maximum particle density is found to drift westward by about 0.4$^{\circ}$ per year, consistently with other measurements. 

The \sax/PM dataset over which this paper is based is available from the authors, upon request.

\begin{acknowledgements}
\sax\ was a joint program of the Italian (ASI) and Dutch (NIVR) space agencies.
\end{acknowledgements}

\bibliographystyle{spphys}      
\bibliography{sax_bkg} 

\end{document}